\documentclass[prd,aps,preprint,showpacs]{revtex4}
\usepackage{epsfig}
\begin{document}

\begin{titlepage}

\voffset 1.5cm

\preprint{
KIAS--P03051,  \hspace{1ex}
UCCHEP/23-03,  \hspace{1ex}
hep-ph/0307385
}

\title{\bf Neutrino Masses, Baryogenesis and Bilinear
R-parity Violation
}

\author{ 
A.G. Akeroyd$^{\mbox{1}}$\footnote{akeroyd@kias.re.kr},
Eung Jin Chun$^{\mbox{1}}$\footnote{ejchun@kias.re.kr},
M.A. D\'\i az$^{\mbox{2}}$\footnote{mad@susy.fis.puc.cl},
Dong-Won Jung$^{\mbox{3}}$}

\affiliation{
$^{1}$Korea Institute for Advanced Study,
207-43 Cheongryangri 2-dong, Dongdaemun-gu,
Seoul 130-722, Republic of Korea \\
$^{2}$Departamento de F\'\i sica, Universidad Cat\'olica de Chile,
Avenida Vicu\~na Mackenna 4860, Santiago, Chile \\
$^{3}$School of Physics, Seoul National University, Seoul
151-747, Republic of Korea
}

\date{July 2003}

\begin{abstract}
\noindent{ We consider the impact of cosmological $B-L$
constraints on supersymmetric standard models with bilinear
breaking of R-parity. In order to avoid erasing any primordial
baryon or lepton asymmetry above the electroweak scale, $B-L$
violation for at least one generation  should be sufficiently small.
Working in the context of models with non--universal soft supersymmetry
breaking masses, we show how the above cosmological constraint can be
satisfied while simultaneously providing a neutrino mass matrix
required by current data.
}

\end{abstract}


\maketitle
\end{titlepage}

\voffset 0cm

In recent years the increasingly strong evidence for neutrino
oscillations from various experiments \cite{Fukuda:1998mi}
has led to the active study of R-parity violating extensions of the
minimal supersymmetric standard model (MSSM)  \cite{Hall:1983id}.
Such models maintain the particle spectrum of the MSSM but
contain renormalizable lepton flavour violating couplings.
The observed neutrino oscillations and mass differences
\cite{Gonzalez-Garcia:2002dz} can be
accommodated with such couplings \cite{hemp,TRpV},
and so these models provide a conceivable
alternative to seesaw mechanisms \cite{seesaw} of neutrino mass generation.
In contrast to the R-parity conserving MSSM,
the lightest supersymmetric particle  is unstable
and decays in the detector with branching ratios  which are correlated
with the neutrino mixing \cite{Mukhopadhyaya:1998xj}.
This provides a robust, experimentally accessible test of the model at the
Large Hadron Collider and/or a $e^+e^-$ Linear Collider \cite{Diaz:2002ij}.
Analogous confirmatory signatures are less readily found
for the elegant seesaw mechanism \cite{seesaw}.
Bilinear R-parity violation (BRpV) is the minimal extension of the MSSM
with R-parity violating terms \cite{hemp,Akeroyd:1997iq,Frank:gs,Banks:1995by}.
The minimal supergravity version of BRpV \cite{Diaz:1997xc}
(i.e. imposing universal soft supersymmetry breaking masses
at an ultraviolet scale)
can easily accommodate the atmospheric neutrino oscillation data.
However, in order to provide the currently favoured large mixing angle
solution for the solar neutrino anomaly, this universality
condition must be relaxed \cite{Chun:1999bq,
Takayama:1999pc,Chun:2002vp}.
Another option for obtaining a realistic neutrino mass matrix is
to allow  both bilinear and trilinear couplings while
keeping the universality condition of the soft supersymmetry breaking masses.
The minimal  model of trilinear R-parity violation (TRpV) assumes
the dominance of the third generation trilinear couplings and thus
contains five free parameters of lepton number violation to fit
all the neutrino data successfully \cite{TRpV}.

The theoretical background on massive Majorana neutrinos and lepton
violating mixing matrices describing neutrino oscillations can be found in
\cite{Schechter:1980gr}.
The atmospheric neutrino data is explained by oscillations
$\nu_{\mu}\longleftrightarrow\nu_{\tau}$, and a global analysis gives the
following $3\sigma$ ranges \cite{Maltoni:2002ni}
\begin{eqnarray}
\label{t23d23range}
0.3 \le& \sin^2\theta_{\mathrm{atm}} &\le 0.7 \cr
1.2 \times 10^{-3}~{\rm{eV}^2} \le& {\Delta m^2_{\mathrm{atm}}}
&\le 4.8 \times 10^{-3}~{\rm{eV}^2}
\end{eqnarray}
with maximal mixing $\sin^2\theta_{\mathrm{atm}}=0.5$ and
${\Delta m^2_{\mathrm{atm}}}=2.5\times 10^{-3}~{\rm{eV}^2}$
as the best fit point. Similarly, the solar neutrino
data is explained by $\nu_e$ oscillation into a mixture of $\nu_{\mu}$
and $\nu_{\tau}$. Global analysis suggests a large mixing angle, although
not maximal, and a much smaller mass squared difference. The allowed
region for ${\Delta m^2_{\mathrm{sol}}}$ previous to KAMLAND results
\cite{Fukuda:1998mi} is now split into two sub-regions. At $3\sigma$
we have \cite{Maltoni:2002ni,Gonzalez-Garcia:2003qf}
\begin{eqnarray}
\label{sol.range}
0.29 \leq& \tan^2\theta_{\mathrm{sol}} &\leq 0.86 \cr
5.1\times 10^{-5}~{\rm{eV}^2} \leq& {\Delta m^2_{\mathrm{sol}}}
&\leq 9.7\times 10^{-5}~{\rm{eV}^2} \cr
1.2\times 10^{-4}~{\rm{eV}^2} \leq& {\Delta m^2_{\mathrm{sol}}}
&\leq 1.9\times 10^{-4}~{\rm{eV}^2}
\end{eqnarray}
with $\tan^2\theta_{\mathrm{sol}}=0.46$ and
${\Delta m^2_{\mathrm{sol}}}=6.9\times 10^{-5}~{\rm{eV}^2}$ as the best fit
point.

In connection with neutrino physics, there appears an important
cosmological consideration. As is well known,  the seesaw
mechanism provides a natural way to generate the baryon asymmetry of the
universe through the out-of-equilibrium decay of a heavy
right-handed neutrino \cite{Fukugita:1986hr}.
Being a new physics model just
around TeV scale, the R-parity violating MSSM can hardly
accommodate such a mechanism of baryogenesis.
However,  in the supersymmetric model, the
so-called Affleck-Dine mechanism  can successfully work to
generate the required amount of the baryon asymmetry  in the flat
direction along, e.g., $L H_u$ \cite{Affleck:1984fy}.  It is notable that such
a property is unaltered even with the presence of R-parity
violating terms which must be very small to generate tiny neutrino
masses.

It is known that lepton number violating couplings have important
consequences for baryogenesis
since together with $B+L$ violating sphaleron processes they are capable
of erasing any pre-existing baryon/lepton asymmetry in the universe
\cite{Fukugita:1990gb, Campbell:1990fa,Dreiner:vm}.
The purpose of this paper is to explicitly check
if such cosmological constraints on the lepton violating
couplings can be satisfied in BRpV while simultaneously
accommodating the form of the neutrino mass matrix
indicated by the atmospheric, solar and reactor neutrino experiments.
A previous analysis \cite{Davidson:1997mc}
derived the cosmological bounds for BRpV but their effect on
the neutrino mass matrix was not covered.
Given the wealth of new data which has
become available since \cite{Davidson:1997mc} appeared, we develop their
analysis and apply the bounds to the currently favoured bimaximal mixing
form of the neutrino mass matrix.

We note that our investigation is not relevant
if electroweak baryogenesis \cite{Kuzmin:1985mm} is operative,
in which case the produced baryon asymmetry cannot be erased
solely by R-parity violating processes. For our purposes we assume that
a $B-L$ asymmetry was generated primordially by some means
at a high energy scale, and our intention is its preservation
at all energies down to the electroweak scale when the sphalerons
finally fall out of equilibrium.


\medskip

We briefly summarize the mechanism of neutrino mass and mixing
generation by R-parity violating couplings,
both bilinear and trilinear. The R-parity violating MSSM
predicts a hierarchical neutrino mass spectrum. The atmospheric
mass scale corresponds approximately to the heaviest
neutrino mass, $m_3$, and it is generated at tree level via a low energy
see-saw mechanism due to the mixing of the neutrinos with the
neutralinos. On the other hand, the solar mass scale,
corresponds approximately to the second heaviest neutrino, $m_2$, and
is generated at the one loop level . The atmospheric neutrino mixing
is also predicted by tree level physics, and depends in a simple way on
sneutrino vacuum expectation values expressed in the basis where the bilinear
parameters are removed from the superpotential. On the other hand,
the solar neutrino mixing angle is again predicted by one-loop
physics which is mainly determined either by the trilinear
couplings in the superpotential or  by the bilinear parameters in
the scalar potential.

Let us remark, however, that we cannot exclude the possibility of
the loop mass dominating over the tree mass, which may have
an interesting implication to  baryogenesis  as
will be discussed later.

\medskip

The well-known baryogenesis constraint \cite{Fukugita:1990gb,Campbell:1990fa}
can be easily applied to the TRpV model with the universality to exclude
this possibility. To see this, let us consider the following trilinear
R-parity violating couplings in the superpotential;
\begin{equation} \label{WRp}
W= \lambda_{ijk} L_i L_j E^c_k + \lambda'_{ijk} L_i Q_j D^c_k
\end{equation}
which generates a neutrino mass at one-loop level as follows;
\begin{equation} \label{Mloop}
M^{loop}_{ij}= 3{\lambda'_{i33}\lambda'_{j33}\over8\pi^2}
{m_b^2(A_b+\mu\tan\beta) \over
m_{\tilde{b}_1}^2-m_{\tilde{b}_2}^2} \ln{m_{\tilde{b}_1}^2\over
m_{\tilde{b}_2}^2} + {\lambda_{i33}\lambda_{j33}\over8\pi^2}
 {m_\tau^2(A_\tau+\mu\tan\beta) \over
m_{\tilde{\tau}_1}^2-m_{\tilde{\tau}_2}^2}
\ln{m_{\tilde{\tau}_1}^2\over m_{\tilde{\tau}_2}^2}\,.
\end{equation}
Note that we have picked up  $\lambda'_{i33}$ and $\lambda_{i33}$
which give the largest contribution to the neutrino masses when
all the trilinear couplings are of similar magnitude.
Then, requiring
the above one-loop mass (\ref{Mloop}) gives rise to the solar
neutrino mass scale, $m_{2} \approx \sqrt{\Delta m^2_\mathrm{sol}}
\approx 8\times 10^{-3}$ eV, we obtain
\begin{equation} \label{formass}
\lambda'_{i33},\; \lambda_{i33}/\sqrt{3} \approx 5\times10^{-5} \left( \tilde{m} \over
300\mbox{ GeV} \right)^{1/2} \left(m_2 \over 8 \mbox{
meV} \right)^{1/2}
\end{equation}
taking
$\tilde{m}=A_b+\mu\tan\beta=m_{\tilde{b}_{1,2}}
=A_\tau+\mu\tan\beta=m_{\tilde{\tau}_{1,2}}$.
Now, the problem is that  such a large coupling makes lepton number
violating interactions very active when the $B+L$ violating
sphaleron interaction is also in thermal equilibrium, so
together they erase the baryon asymmetry before the electroweak
phase transition.  Indeed, the interaction in Eq.~(\ref{WRp})
gives the decay width for lepton number violating one-to-two body
decays,
\begin{equation} \label{Gamma12}
 \Gamma_{12} = {\pi \lambda^{(\prime)2}_{i33} \over 192 \zeta(3)}
 {\tilde{m}^2 \over T}
\end{equation}
assuming $T\gg \tilde{m}$.  The out-of-equilibrium condition,
$\Gamma_{12} < H=1.66 \sqrt{g_{eff}} T^2/m_{Pl}$, gives
\begin{equation} \label{forbaryo}
 \lambda'_{i33},\; \lambda_{i33} < 2\times10^{-7} \left( \tilde{m} \over 300\mbox{
 GeV} \right)^{1/2}
\end{equation}
for $g_{eff}=915/4$.  This is for $T >> \tilde{m}$. An improved result which does
not make this assumption was presented in \cite{Dreiner:vm} and shows that the
$T/\tilde{m}$ dependence of Eq.~(\ref{Gamma12}) is very mild.
One sees a big contradiction between
(\ref{formass}) and (\ref{forbaryo}).
As indicated in Eq.~(\ref{formass}), one needs the trilinear couplings of
$\lambda'_{233,333} \sim \lambda_{133,233}\sim {\cal O}(10^{-5})$ to accommodate
the required bi-large mixing of the atmospheric and
solar neutrino oscillation \cite{TRpV}.
Thus, the baryogenesis constraint rules out a purely TRpV explanation of
the observed neutrino data.

\medskip

The situation may be different if the neutrino masses are
generated purely by bilinear R-parity violating couplings with non-universal
soft masses, in which case the non-universality can give much freedom.
Forbidding the lepton number violating trilinear couplings
in the superpotential in Eq.~(\ref{WRp}),
the BRpV model allows the following
dimension-two terms in the superpotential and in the soft
supersymmetry breaking scalar potential:
\begin{eqnarray} \label{biRp}
 W &=& \mu (\epsilon_i L_i H_2 + H_1 H_2 )
 \nonumber\\
 V_{soft} &=& \mu(\epsilon_i B_i L_i H_2 + B H_1 H_2) + m^2_{L_i
 H_1} L_i H_1^\dagger + h.c.
\end{eqnarray}
Here we have used the same notation for the superfields and their
scalar components.    A key point to notice is that without the
electroweak symmetry breaking, the $SU(4)$ rotation in the
`superfields', $L_i$ and $H_1$;
\begin{equation} \label{su4}
L_i \to L_i +\epsilon_i H_1\quad\mbox{and}\quad
 H_1 \to H_1 -\epsilon_i L_i
\end{equation}
which gets rid of  the $\epsilon_i$ term (valid up to
$O(\epsilon_i)$) leaves invariant the gauge interactions and thus
its effect is only to generate the effective couplings as in
Eq.~(\ref{WRp}) with
\begin{equation} \label{efflam}
 \lambda'_{i33} = \epsilon_i h_b \quad\mbox{and}\quad
 \lambda_{i33} = \epsilon_i h_\tau \,.
\end{equation}
Under the $SU(4)$ rotation (\ref{su4}), the scalar potential in
(\ref{biRp}) becomes
\begin{equation} \label{Vsoft}
 V_{soft}= \mu (B H_1 H_2 - \epsilon_i \Delta B_i L_i H_2)
         + (m_{L_i H_1}^2 -\epsilon_i \Delta m^2_i) L_i
         H_1^\dagger + h.c.
 \end{equation}
where $\Delta B_i = B-B_i$ and $\Delta m^2_i =
m_{H_1}^2-m_{L_i}^2$.   Eq.~(\ref{Vsoft}) shows that the
additional lepton number violating mixing mass terms for the
`scalar fields' $\tilde{L}_i$ and $H_{1,2}$ (in the basis of
vanishing $\epsilon_i$)  arise in the presence of the
non-universal soft supersymmetry breaking parameters.
Diagonalizing away such mixing mass terms can be made by the
following  rotation among the scalar fields $\tilde{L}_i$, $H_1$
and $H'_2 \equiv i\tau_2 H_2^\dagger$:
\begin{eqnarray} \label{LHrot}
&&\tilde{L}_i \to \tilde{L}_i - \varepsilon_{i1} H_1 -
\varepsilon_{i2} H'_2
\nonumber\\
&&H_1 \to H_1 +\varepsilon_{i1} \tilde{L}_i
\nonumber\\
&&H'_2 \to H'_2 + \varepsilon_{i2} \tilde{L}_i
\end{eqnarray}
where the variables $ \varepsilon_{i1}$ and  $\varepsilon_{i2}$
are determined as
\begin{eqnarray} \label{varepses}
 \varepsilon_{i1} &=& {
(m^2_{H_2}+\mu^2-m_{L_i}^2)(\epsilon_i\Delta m^2_i-m^2_{L_iH_1})
-\epsilon_i\mu^2 B \Delta B_i \over
(m^2_{H_1}+\mu^2-m^2_{L_i})(m^2_{H_2}+\mu^2-m^2_{L_i})-\mu^2
B^2 } \nonumber\\
\varepsilon_{i2} &=& { (m^2_{H_1}+\mu^2-m_{L_i}^2)\epsilon_i
\mu\Delta B_i - \mu B (\epsilon_i\Delta m^2_i-m^2_{L_iH_1}) \over
(m^2_{H_1}+\mu^2-m^2_{L_i})(m^2_{H_2}+\mu^2-m^2_{L_i})-\mu^2 B^2 }
\end{eqnarray}
As will be discussed later, it is useful
to rewrite $\varepsilon_{i1,i2}$ in terms of the variables
$\xi_i$ and $\eta_i$
defined by
$$
 \xi_i \equiv{\langle \tilde{\nu}_i \rangle \over \langle H_1 \rangle}
 -\epsilon_i \quad\mbox{and}\quad
 \eta_i \equiv \xi_i + \epsilon_i{\Delta B_i \over B}
$$
where $\langle \tilde{\nu}_i \rangle$ and  $\langle H_1 \rangle$ are
the vacuum expectation values of the sneutrino and Higgs boson
generated after the electroweak symmetry breaking.
Using the  minimization condition of the Higgs and sneutrino fields,
we obtain
\begin{eqnarray} \label{vareps2}
\varepsilon_{i1} &=&  -\xi_i -\eta_i {m_A^2 s_\beta^2
(m^2_{\tilde{\nu}_i} - M^2_Z c_{2\beta})  \over
m^2_{\tilde{\nu}_i} (m^2_{\tilde{\nu}_i}-m^2_A)
-(m^2_{\tilde{\nu}_i}- m_A^2 s_\beta^2)M_Z^2 c_{2\beta}  }
\nonumber\\
\varepsilon_{i2} &=& {\eta_i \over t_\beta} { m_A^2 s_\beta^2
m^2_{\tilde{\nu}_i} \over m^2_{\tilde{\nu}_i}
(m^2_{\tilde{\nu}_i}-m^2_A) -(m^2_{\tilde{\nu}_i}- m_A^2
s_\beta^2)M_Z^2 c_{2\beta} }
\end{eqnarray}
where $t_\beta = \tan\beta = \langle H_2 \rangle/  \langle H_1 \rangle$.
The variables $\varepsilon_{i1,i2}$ control the size of lepton
number violating interactions which now arise due to the
misalignment between the scalars, $\tilde{L}_i$ and $H_{1,2}$, and
fermions, $L_i$ and $\tilde{H}_{1,2}$.   That is,  the rotation
(\ref{LHrot}) gives rise to  the following  lepton number
violating vertices:
\begin{eqnarray} \label{effL}
{\cal L}_{eff} &=& h_\tau\varepsilon_{i1} \tilde{L}_i L_3 E^c_3 +
h_b\varepsilon_{i1} \tilde{L}_i Q_3 D^c_3 +
h_t\varepsilon_{i2}\tilde{L}'_i Q_3 U^c_3
\nonumber\\
&& + {g'\varepsilon_{i1}\over\sqrt{2}} [ H_1^\dagger L_i \tilde{B}
+ \tilde{L}_i^\dagger \tilde{H}_1 \tilde{B}] +
{g'\varepsilon_{i2}\over\sqrt{2}}\tilde{L}_i^\dagger \tilde{H}'_2
\tilde{B}
\nonumber\\
&& + {g\varepsilon_{i1}\over\sqrt{2}} [ H_1^\dagger \tau^a L_i
\lambda^a  + \tilde{L}_i^\dagger \tau^a\tilde{H}_1 \lambda^a] +
{g\varepsilon_{i2}\over\sqrt{2}}\tilde{L}_i^\dagger \tau^a
\tilde{H}'_2 \lambda^a
 + h.c.
\end{eqnarray}
where $\tilde{L}'_i \equiv i\tau_2 \tilde{L}_i^\dagger$, $\tau^a$
are Pauli matrices and $\lambda^a$ represent the $SU(2)$ gauginos.
Applying the constraint (\ref{forbaryo}) to the couplings in
Eqs.~(\ref{efflam}) and (\ref{effL}), we get \cite{Davidson:1997mc}

\begin{eqnarray} \label{forbaryo2}
 \epsilon_i
  &<& 1.2\times 10^{-5}c_\beta \left( \tilde{m} \over
 300 \mbox{ GeV} \right)^{1/2} \nonumber\\
 \varepsilon_{i1}
  &<& 3\times 10^{-7} \left( m_{\chi^0} \over
 300 \mbox{ GeV} \right)^{1/2} \\
 \varepsilon_{i2} &<& 2\times 10^{-7} s_\beta \left( m_{L_i} \over
 300 \mbox{ GeV} \right)^{1/2}
 \nonumber
\end{eqnarray}
where $\tilde{m}$ is the smallest mass of the sfermions involved
in the $\lambda'_{i33}$ term; $L_i, Q_3$ and $D^c_3$, $m_{\chi}$
is a gaugino mass involved in the process $\chi \to L_i H_1$ and
the  last equation comes from the process $\tilde{L_i} \to Q_3
U^c_3$.

\medskip

The sizes of certain bilinear parameters are determined
to generate realistic neutrino masses and mixing in our bilinear model.
First of all, upon electroweak symmetry breaking,
the Higgs and sneutrino acquire vacuum expectation values and
generate a tree-level neutrino mass matrix
\begin{equation} \label{Mtree}
 M^{tree}_{ij} = {M_Z^2 \over F_N} \xi_i \xi_j c_\beta^2
\end{equation}
where $F_N= M_1M_2/(c_W^2 M_1+ s_W^2 M_2)+ M_Z^2 c_{2\beta}/\mu$
\cite{Chun:1999bq}.   Recall that $\xi_i$ arises through the
mismatch of soft terms between $L_i$ and $H_1$ as follows;
\begin{equation} \label{xiis}
 \xi_i = \epsilon_i {\Delta m_i^2 + \Delta B_i \mu t_\beta-
  m^2_{L_i H_1}/\epsilon_i \over m^2_{\tilde{\nu}_i} } \,.
\end{equation}
The tree mass in Eq.~(\ref{Mtree}) gives the heavier mass scale,
$m_{3}={M_Z^2 \over F_N} \xi^2 c_\beta^2$.  Considering the
atmospheric neutrino mass-squared difference, $\Delta m^2_\mathrm{atm}
\approx 2.5\times10^{-3}$ eV $\approx m_{3}^2$, we get
\begin{equation}
 \xi c_\beta = 7.4\times 10^{-7} \left(F_N \over M_Z\right)^{1/2}
  \left( m_{3} \over 0.05 \mbox{ eV} \right)^{1/2}
\end{equation}
Since the two mixing angles,  $\theta_{23}=\theta_\mathrm{atm}$
and  $\theta_{13}$, satisfy
\begin{equation}
\tan\theta_{23}=\xi_2/\xi_3\approx 1\,,
\qquad
|\tan\theta_{13}|=|\xi_1|/\sqrt{\xi_2^2+\xi_3^2}
\ll 1
\label{tlangles}
\end{equation}
we need $\xi_1 < 0.3 \xi_{2,3}$ to make small $\theta_{13}$ and
$\xi_2 \approx \xi_3$ for near maximal atmospheric mixing. Thus,
current neutrino oscillation data require
\begin{equation} \label{xisize}
 \xi_1 \ll \xi_2 \approx \xi_3 \approx 5.2\times 10^{-7}
 {1\over c_\beta}  \left(F_N \over M_Z\right)^{1/2}
  \left( m_3 \over 0.05 \mbox{ eV} \right)^{1/2}
\end{equation}
Let us now consider how one-loop corrections generate the
neutrino masses and mixing accounting for the solar neutrino
oscillation.
In the bilinear model, the bi-large mixing of the
atmospheric and solar neutrinos cannot be obtained under the
assumption of universal soft terms \cite{Chun:1999bq}.  Thus, one needs
to introduce non-universality in soft terms in order  to
accommodate the large solar mixing.

\medskip

Depending on the degrees of
the deviation from the universality, we can consider two cases. First,
the non-universality of soft parameters can arise due to small
mismatches (likely to be caused by some threshold corrections) in
the renormalization group evolution. In this case, the quantities
$\Delta m^2_i$, $m^2_{L_i H_1}/\epsilon_i$ and $\mu \Delta B_i$
are much smaller than the typical soft mass-squared $\tilde{m}^2$
so that the induced trilinear couplings in Eq.~(\ref{efflam}) give
the major contribution to  the size of
$m_{2}\approx \sqrt{\Delta m^2_\mathrm{sol}}$
\cite{Takayama:1999pc,Diaz:2003as}.
As discussed before, this causes the contradiction of
Eqs.~(\ref{formass}) and (\ref{forbaryo}). In other words, the
condition of $m_{2}\sim 8$ meV yields $\epsilon_i \sim
4\times10^{-3} c_\beta$ which is far above the first constraint in
Eq.~(\ref{forbaryo2}).

However, we point out that there is a different way of reconciling the
neutrino data with the baryogenesis requirement.
Note that one cannot exclude the possibility
that the loop mass is larger than the tree mass. For instance,
one can take the superpotential bilinear parameter $\epsilon_i$
much larger than $\xi_i$, accepting a very small deviation of
the non-universality or a cancellation among the terms \cite{TRpV},
see Eq.~(\ref{xiis}).
In this situation, the heavier
neutrino mass scale can be produced mainly by the bottom-sbottom loop which
can be rewritten from  Eqs.~(\ref{Mloop}) and (\ref{efflam}) as follows:
\begin{equation}
M^{loop}_{ij}= {3 h_b^2\over8\pi^2} \epsilon_i \epsilon_j
{m_b^2(A_b+\mu\tan\beta) \over
m_{\tilde{b}_1}^2-m_{\tilde{b}_2}^2} \ln{m_{\tilde{b}_1}^2\over
m_{\tilde{b}_2}^2} \,.
\end{equation}
As  the above loop contribution determines the atmospheric neutrino
mass and mixing, the condition (\ref{xisize}) has to be replaced by
\begin{equation}
 \epsilon_1 \ll \epsilon_2 \approx \epsilon_3 \approx
 8\times10^{-3} c_\beta \left( \tilde{m}\over 300 \mbox{ GeV} \right)^{1/2}.
\end{equation}
Similarly to the previous discussions,
$\epsilon_{2,3}$ cannot satisfy the baryogenesis constraint
(\ref{forbaryo2}) at all.  But, $\epsilon_1$ can be made arbitrarily small.
Let us recall that it is sufficient to suppress lepton number
violating couplings for just one lepton flavour.  In our case, it is
the electron number, which is implied by the smallness of $\theta_{13}$.
Now, in order for the tree mass (\ref{Mtree}) to produce the solar
neutrino mass and mixing, we need
\begin{equation}
 \xi_1 \sim \xi_2 \sim 3\times10^{-7}{1\over c_\beta}
 \left( F_N \over M_Z \right)^{1/2}\,.
\end{equation}
For such  small $\epsilon_1$ and the small deviation of the universality,
we expect $\xi_1 \simeq \eta_1$ and thus the variables
$\varepsilon_{11,12}$ in Eq.~(\ref{vareps2}) can be approximated by
\begin{equation}
 \varepsilon_{11} \approx -\xi_1 {m^2_{\tilde{\nu}_1}-m^2_A c_\beta^2
  \over m^2_{\tilde{\nu}_1}- m_A^2}\,,\quad
 \varepsilon_{12} \approx -{\xi_1 \over t_\beta}
 {m_A^2  s^2_\beta \over m^2_{\tilde{\nu}_1}- m_A^2} \,,
\end{equation}
neglecting $M_Z^2$ terms.  From this, one can see that
the baryogenesis constraint (\ref{forbaryo2})
can be  satisfied {\em if}
$ 1 < t_\beta <  (m^2_{\tilde{\nu}_1}- m_A^2)/
(m^2_{\tilde{\nu}_1}-m^2_A c_\beta^2)$.

\medskip

Secondly, we consider the more general non-universality
implying that  $\Delta m^2_i$, $m^2_{L_i
H_1}/\epsilon_i$ and $\mu \Delta B_i$ are of the order
$\tilde{m}^2$.  In this case, the neutral scalar and neutralino
exchange loops can give important contributions to the one-loop mass
as long as $\tan\beta$ is not
too large and the large misalignment between $\xi_i$ and $\eta_i$ is allowed.
Adopting the result of Ref.~\cite{Chun:2002vp},
the one-loop mass coming from the neutral scalar loops
is roughly given by
\begin{equation} \label{Msloop}
M^{loop}_{ij} \approx {g^2\over64\pi^2} m_{\chi^0}
\theta_{i\phi}\theta_{j\phi} B_0(m^2_{\chi^0}, m_\phi^2)
\end{equation}
where $B_0(x,y)=-{x\over x-y} \ln{x\over y} -\ln{x\over Q^2} +1$
and $\phi$ represents the neural Higgs bosons, $h, H$ and $A$.
Neglecting unimportant contribution of $\xi_i$, the variables
$\theta_{i\phi}$ are approximately given by
\begin{eqnarray} \label{thetas}
 \theta_{ih} &\approx&  \eta_i s_\beta m_A^2
    { m_{\tilde{\nu}_i}^2 c_{\alpha-\beta}
    - M_Z^2 c_{2\beta} c_{\alpha+\beta} \over
      (m_{\tilde{\nu}_i}^2-m_h^2) (m_{\tilde{\nu}_i}^2-m_H^2) }
      \nonumber\\
 \theta_{iH} &\approx&  \eta_i s_\beta m_A^2
    { m_{\tilde{\nu}_i}^2  s_{\alpha-\beta}
    - M_Z^2  c_{2\beta} s_{\alpha+\beta}  \over
      (m_{\tilde{\nu}_i}^2-m_h^2) (m_{\tilde{\nu}_i}^2-m_H^2) }
         \nonumber\\
 \theta_{iA} &\approx&
         i \eta_i s_\beta { m_A^2 \over m_A^2- m_{\tilde{\nu}_i}^2}
\end{eqnarray}
where 
$m_{h,H}$ are
the Higgs boson masses at tree-level determined  by
$m^2_{h,H}=1/2[m^2_A+M_Z^2 \mp \sqrt{ (m^2_A+M_Z^2)^2-4
m_A^2 M_Z^2 c_{2\beta}^2} ]$, and the angle $\alpha$ is defined by
$c_{2\alpha}=c_{2\beta} (m_A^2-M_Z^2)/(m_h^2-m_H^2)$ and
$s_{2\alpha}=s_{2\beta} (m_A^2+M_Z^2)/(m_h^2-m_H^2)$.
Our convention for the pseudo-scalar Higgs boson mass is
that $m_A^2 = -\mu B/c_\beta s_\beta$.
Requiring $m_{2} \sim M^{loop}_{ij}$, one obtains
\begin{equation} \label{thetasize}
\theta_{i\phi} \sim 6\times10^{-6} \left( 300\mbox{ GeV}\over
m_{\chi^0} \right)^{1/2} \left(m_\phi \over m_{\chi^0}\right)
\left( m_{2} \over 8 \mbox{ meV} \right)
\end{equation}
As discussed in Ref.~\cite{Chun:2002vp}, the large mixing
of solar neutrinos require $\theta_{1\phi} \sim \theta_{2\phi}$.
This has to be contrasted the condition $\xi_2\approx \xi_3$ (\ref{xisize})
for the large atmospheric neutrino mixing.

\medskip

{}From Eqs.~(\ref{forbaryo2}), (\ref{xisize}) and (\ref{thetasize}),
one sees that the couplings $\varepsilon_{i1,i2}$ are required to be
smaller than $\xi_i$ or $\theta_{i\phi}$ by one-order of magnitude.
Thus, it is generally difficult to satisfy both the baryogenesis
constraints and obtain the realistic neutrino masses and mixing.
However, it is not impossible to find some reasonable parameter space
where both requirements are reconciled, which is due to the fact that
the variables $\varepsilon_{i1,i2}$ and $\xi_i$ or $\theta_{i\phi}$
have different dependencies on the input parameters.
Comparing Eq.~(\ref{vareps2}) with Eq.~(\ref{thetas}), we notice
that $\varepsilon_{i1,i2}$ (or $\xi_i$ and $\eta_i$)
can be made small while keeping
$\theta_{i\phi} \sim 6\times10^{-6}$ (\ref{thetasize})
when the sneutrino mass $m_{\tilde{\nu}_i}$ is close to one
of the Higgs boson masses, $m_h, m_H$ and $m_A$.
Since the heavy Higgs scalar mass,
$m_H$, is usually close to the pseudo scalar mass, $m_A$,
and $s^2_\beta\approx 1$, we
find it better that the sneutrino mass is closer to the light
Higgs scalar mass, that is, $m_{\tilde{\nu}_1} \sim m_h$.
Barring cancellation, both terms in $\varepsilon_{i1}$ (\ref{vareps2})
should be less than $3\times 10^{-7}$.  Again here,
this is possible for $i=1$, that is,
the electron number violating parameters,
$\varepsilon_{11, 12}$, can only be suppressed for our purpose.
For illustration,
let us calculate $\varepsilon_{i1}+\xi_i$, $\varepsilon_{i2}$
and $\theta_{i\phi}$ for the cases with $m_A=100,300$ GeV and
$\tan\beta=3,30$.  In what follows, we present the values of $(
\theta_{ih}, \theta_{iH},\theta_{iA}; \varepsilon_{i1}+\xi_i,
\varepsilon_{i2})$ normalized with $\varepsilon_{i2}=1$,
indicating the rough ranges of $m_{\tilde{\nu}_i}$ allowing
for $ \varepsilon_{i1} > \theta_{i\phi}/20$:
\begin{eqnarray}
\mbox{Case 1} \label{case1} &&
\qquad t_\beta=3,\; m_A=100 \mbox{ GeV}, \;  m_h=60\mbox{ GeV} \\
&& (+77, +64, -9.1; -3.6, 1)
\quad\mbox{for}\quad m_{\tilde{\nu}_i}=55 \mbox{ GeV} \nonumber\\
&& (-22, -21, -6.5; -0.96, 1) \quad\mbox{for}\quad
m_{\tilde{\nu}_i}=71 \mbox{ GeV} \nonumber \\
 \mbox{Case 2} \label{case2} &&
\qquad t_\beta=30,\; m_A=100 \mbox{ GeV}, \;  m_h=90\mbox{ GeV} \\
&& (-113,+342,-77; -17,~1)
 \quad\mbox{for}\quad m_{\tilde{\nu}_i}=73 \mbox{ GeV} \nonumber\\
&& (-66, -210, -49; +11, ~1)
 \quad\mbox{for}\quad m_{\tilde{\nu}_i}=115 \mbox{ GeV} \nonumber\\
\mbox{Case 3} \label{case3} &&
\qquad t_\beta=3,\; m_A=300 \mbox{ GeV}, \;  m_h=72\mbox{ GeV} \\
&& (+29, +53, -8.4; -1.5,~ 1)
 \quad\mbox{for}\quad m_{\tilde{\nu}_i}=60\mbox{ GeV} \nonumber\\
&& (-10, -26, -5.5; +0.54, ~1)
 \quad\mbox{for}\quad m_{\tilde{\nu}_i}=90 \mbox{ GeV} \nonumber\\
\mbox{Case 4} \label{case4} &&
\qquad t_\beta=30,\; m_A=300 \mbox{ GeV}, \;  m_h=91\mbox{ GeV} \\
&& (+18, +289, -74; -14, ~1)
 \quad\mbox{for}\quad m_{\tilde{\nu}_i}=75\mbox{ GeV} \nonumber\\
&& (-6.9, -212, -49; +11, ~1)
 \quad\mbox{for}\quad m_{\tilde{\nu}_i}=115\mbox{ GeV} \nonumber
\end{eqnarray}
{}From the above calculation, one can see that the non-erasure
condition can be satisfied if the difference between  the
sneutrino and the light Higgs boson mass is within 10\%.
In order to confirm the above properties, we made a numerical
calculation to find a set of points satisfying both
the baryogenesis constraints and the atmospheric and solar
neutrino data.  For this, we incorporate the exact formulae for
the neutrino mass matrix derived in Ref.~\cite{Chun:2002vp}.
In Figures 1 and 2, we plot the variable $\varepsilon_{11}$
in terms of the electron sneutrino mass $m_{\tilde{\nu}_1}$ for
all the points accommodating all the observed neutrino data for
Cases 1 and 2.
The plots clearly show the suppression of $\varepsilon_{11}$
when the sneutrino mass is close to a Higgs boson mass.
Similar behavior is also found in Cases 3 and 4.

Another way of suppressing $\varepsilon_{i1}$ is to arrange
 a cancellation between two terms in $\varepsilon_{i1}$.
{}From Eqs.~(\ref{vareps2}) and (\ref{thetas}), one generally has
\begin{equation}
 \varepsilon_{i1} \sim -\xi_i - t_\beta \varepsilon_{i2}
 \quad\mbox{and}\quad \theta_{i\phi} \sim t_\beta \varepsilon_{i2}
\end{equation}
for $m_{\tilde{\nu}_i} \gg M_Z$.  Now, one can see that the
conditions (\ref{forbaryo2}) and (\ref{thetasize}) can be
satisfied for $t_\beta\sim 30$ with the cancellation in
$\varepsilon_{i1} \sim \xi_i + \theta_{i\phi}$.  Again, this can
work only for the electron direction with $\xi_1\sim
\theta_{1\phi}$ since Eq.~(\ref{xisize}) shows $\xi_{i} \gg
\theta_{i\phi}\sim 6\times10^{-6}$ for $i=2,3$ and large $\tan\beta$.

\medskip

\medskip

In conclusion,  we have investigated how the cosmological requirement
for a successful baryogenesis can be reconciled with a realistic
neutrino mass matrix in the R-parity violating version of
supersymmetric standard model.    Our main focus has been to see
whether the $B-L$ violating interactions can be sufficiently
suppressed in order not to erase  a pre-existing baryon or lepton
asymmetry of the universe.  Such a baryogenesis
constraint cannot be satisfied if the trilinear R-parity violating couplings
are introduced to explain the atmospheric and solar neutrino masses and mixing
under the assumption of the universal soft supersymmetry breaking masses.
In the bilinear model, the observed neutrino data can be well
explained  if the non-universality is  allowed.
Our analysis shows that the non-erasure condition can be met
by suppressing the electron number violating parameters,
which is related to the smallness of the angle $\theta_{13}$.
In the case of a large violation of the universality,
the electron sneutrino mass has to be nearly degenerate
with the light Higgs scalar mass.
For a small violation of the universality,
we argued that the situation of the loop mass dominating
over the tree mass is preferred contrary to the usual consideration.
A consequence of our analysis is that the bilinear R-parity violating
supersymmetric standard model can provide a  framework not only
for a realistic neutrino mass matrix but also for a successful
baryogenesis through the Affleck-Dine Mechanism.
Finally, let us note that our consideration is not relevant
if the electro-weak baryogenesis is operative.

\medskip

{\bf Acknowledgments}: EJC was supported by the Korea Research
Foundation Grant, KRF-2002-070-C00022 and DWJ by KRF-2002-015-CP0060.
MAD was supported by Conicyt grant No.~1030948, and would like to thank
the KIAS High Energy Group for their kind hospitality.
We thank Sacha Davidson for discussions and IPPP for its hospitality.

\newpage

\begin{figure}
\epsfig{file=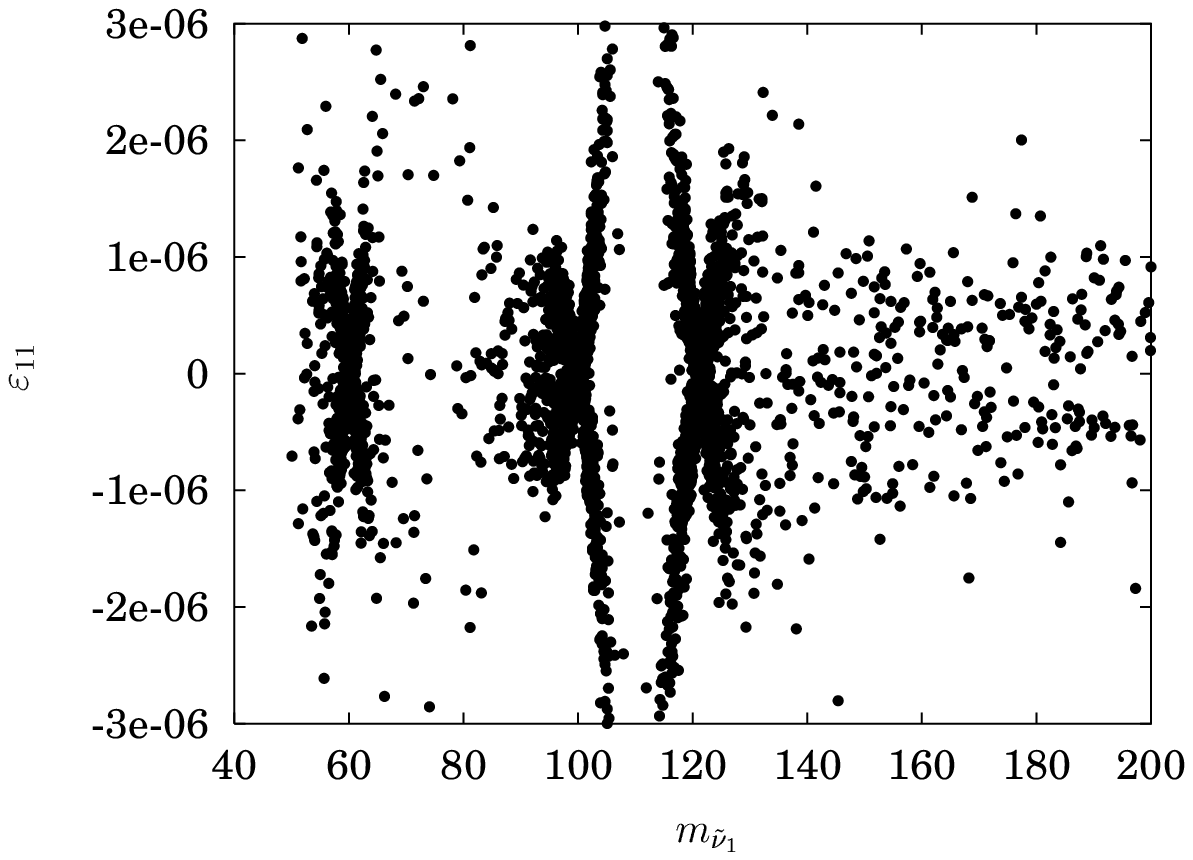,height=8cm,width=13cm}
\caption{ 
The quantity $\varepsilon_{11}$ is shown as a function of the
electron sneutrino mass $m_{\tilde{\nu}_1}$ for all the points
generating the required neutrino masses and mixing for Case 1 in Eq.~(\ref{case1}).
}
\end{figure}

\begin{figure}
\epsfig{file=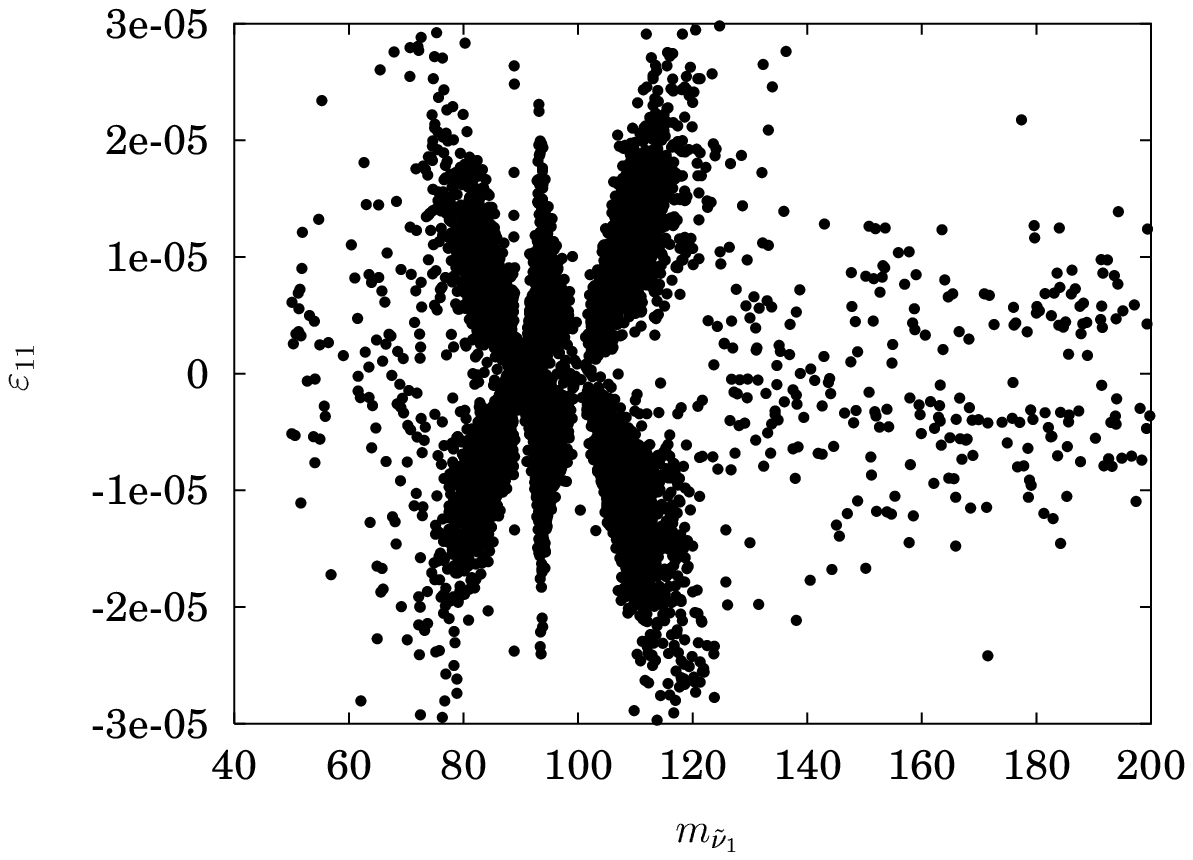,height=8cm,width=13cm}
\caption{ 
Same as FIG.\ 1 for Case 2 in Eq.~(\ref{case2}).
}
\end{figure}


\begin{thebibliography}{99}

\bibitem{Fukuda:1998mi}
Y.~Fukuda {\it et al.}  [Super-Kamiokande Collaboration],
Phys.\ Rev.\ Lett.\  {\bf 81}, 1562 (1998);
M.~Apollonio {\it et al.}  [CHOOZ Collaboration],
Phys.\ Lett.\ B {\bf 420}, 397 (1998);
Q.~R.~Ahmad {\it et al.}  [SNO Collaboration],
Phys.\ Rev.\ Lett.\  {\bf 87}, 071301 (2001);
S.~Fukuda {\it et al.}  [Super-Kamiokande Collaboration],
Phys.\ Lett.\ B {\bf 539}, 179 (2002);
K.~Eguchi {\it et al.}  [KamLAND Collaboration],
Phys.\ Rev.\ Lett.\  {\bf 90}, 021802 (2003);
M.~H.~Ahn {\it et al.}  [K2K Collaboration],
Phys.\ Rev.\ Lett.\  {\bf 90}, 041801 (2003).

\bibitem{Hall:1983id}
L.~J.~Hall and M.~Suzuki,
Nucl.\ Phys.\ B {\bf 231}, 419 (1984).

\bibitem{Gonzalez-Garcia:2002dz}
M.~C.~Gonzalez-Garcia and Y.~Nir,
Rev.\ Mod.\ Phys.\  {\bf 75}, 345 (2003);
S.~Pakvasa and J.~W.~Valle,
arXiv:hep-ph/0301061;
A.~Y.~Smirnov,
Nuovo Cim.\  {\bf 117B} (2002) 1237.

\bibitem{hemp}
R. Hempfling, Nucl.\ Phys.\ {\bf B478}, 3 (1996);
M.~Hirsch {\sl et. al.}, Phys.\ Rev.\ D {\bf 62}, 113008 (2000) [Erratum-ibid.\ D
{\bf 65}, 119901 (2002)].

\bibitem{TRpV}
E.~J.~Chun, S.~K.~Kang, C.~W.~Kim and U.~W.~Lee, Nucl.\ Phys.\ B
{\bf 544}, 89 (1999);
E.~J.~Chun, D.~W.~Jung, S.~K.~Kang and J.~D.~Park,
Phys.\ Rev.\ D {\bf 66}, 073003 (2002).

\bibitem{seesaw}
 T.~Yanagida, in proc. of the {\it Workshop on the Unified Therory and Baryon
  Number in the Universe}, O.Sawada and A.Sugamoto eds., KEK report 79-18,
  1979, p.95, Tsukuba, Japan;
 S.~L. Glashow, in {\it Cargese 1979, Proceedings, Quarks and Leptons}, 687-713;
 M.~Gell-Mann, P.~Ramond, and R.~Slansky, in {\it Supergravity},
  P. van Nieuwenhuizen and D.Z. Freedman ed., North Holland, Amsterdam 1979,
  p.315;
R.~N.~Mohapatra and G.~Senjanovic,
Phys.\ Rev.\ Lett.\  {\bf 44}, 912 (1980).

\bibitem{Mukhopadhyaya:1998xj}
B.~Mukhopadhyaya, S.~Roy and F.~Vissani,
Phys.\ Lett.\ B {\bf 443}, 191 (1998);
E.~J.~Chun and J.~S.~Lee,
Phys.\ Rev.\ D {\bf 60}, 075006 (1999);
S.~Y.~Choi, E.~J.~Chun, S.~K.~Kang and J.~S.~Lee,
Phys.\ Rev.\ D {\bf 60}, 075002 (1999);
W.~Porod, M.~Hirsch, J.~Romao and J.~W.~Valle,
Phys.\ Rev.\ D {\bf 63}, 115004 (2001);
J.~C.~Romao {\sl et. al.},
Phys.\ Rev.\ D {\bf 61} (2000) 071703.

\bibitem{Diaz:2002ij}
F.~De Campos {\sl et. al.},
Nucl.\ Phys.\ B {\bf 623} (2002) 47;
M.~A.~Diaz, R.~A.~Lineros and M.~A.~Rivera,
Phys.\ Rev.\ D {\bf 67}, 115004 (2003).

\bibitem{Akeroyd:1997iq}
A.~G.~Akeroyd {\sl et. al.},
Nucl.\ Phys.\ B {\bf 529} (1998) 3;
M. Nowakowski and A. Pilaftsis, Nucl.\ Phys.\ B {\bf 461}, 19
(1996);
R. Kitano and K. Oda, Phys.\ Rev.\ D {\bf 61}, 113001 (2000);
A.~G.~Akeroyd, M.~A.~Diaz and J.~W.~Valle,
Phys.\ Lett.\ B {\bf 441} (1998) 224;
A. Yu. Smirnov and F. Vissani, Nucl.\ Phys.\ B {\bf 460}, 37 (1996);
J.~M.~Mira, E.~Nardi, D.~A.~Restrepo and J.~W.~Valle,
Phys.\ Lett.\ B {\bf 492} (2000) 81.

\bibitem{Frank:gs}
M.~Frank,
Phys.\ Rev.\ D {\bf 62} (2000) 015006;
M. A. D\'{\i}az, E. Torrente--Lujan and J. W. F. Valle,
Nucl.\ Phys.\ B {\bf 551}, 78 (1999);
B.~de Carlos and P.~L.~White,
Phys.\ Rev.\ D {\bf 54} (1996) 3427;
B.~de Carlos and P.~L.~White,
Phys.\ Rev.\ D {\bf 55} (1997) 4222;
D.~Suematsu,
Phys.\ Lett.\ B {\bf 506} (2001) 131;
J.~Ferrandis,
Phys.\ Rev.\ D {\bf 60} (1999) 095012;
M.~A.~Diaz, J.~Ferrandis and J.~W.~Valle,
Nucl.\ Phys.\ B {\bf 573} (2000) 75.

\bibitem{Banks:1995by}
T.~Banks, Y.~Grossman, E.~Nardi and Y.~Nir,
Phys.\ Rev.\ D {\bf 52} (1995) 5319;
M.~A.~Diaz, J.~Ferrandis, J.~C.~Romao and J.~W.~Valle,
Phys.\ Lett.\ B {\bf 453} (1999) 263, and
Nucl.\ Phys.\ B {\bf 590} (2000) 3;
H.~P.~Nilles and N.~Polonsky,
Nucl.\ Phys.\ B {\bf 484} (1997) 33;
D.~E.~Kaplan and A.~E.~Nelson,
JHEP {\bf 0001} (2000) 033;
M.~A.~Diaz, D.~A.~Restrepo and J.~W.~Valle,
Nucl.\ Phys.\ B {\bf 583} (2000) 182.

\bibitem{Diaz:1997xc}
M.~A.~Diaz, J.~C.~Romao and J.~W.~Valle,
Nucl.\ Phys.\ B {\bf 524}, 23 (1998).

\bibitem{Chun:1999bq}
E.~J.~Chun and S.~K.~Kang,
Phys.\ Rev.\ D {\bf 61}, 075012 (2000).

\bibitem{Takayama:1999pc}
F.~Takayama and M.~Yamaguchi,
Phys.\ Lett.\ B {\bf 476}, 116 (2000);
S.~Davidson and M.~Losada,
JHEP {\bf 0005}, 021 (2000);
A.~Abada, S.~Davidson and M.~Losada,
Phys.\ Rev.\ D {\bf 65}, 075010 (2002);
A.~S.~Joshipura, R.~D.~Vaidya and S.~K.~Vempati,
Nucl.\ Phys.\ B {\bf 639}, 290 (2002).

\bibitem{Chun:2002vp}
E.~J.~Chun, D.~W.~Jung and J.~D.~Park,
Phys.\ Lett.\ B {\bf 557}, 233 (2003).

\bibitem{Diaz:2003as}
M.~A.~Diaz, M.~Hirsch, W.~Porod, J.~C.~Romao and J.~W.~Valle,
arXiv:hep-ph/0302021.

\bibitem{Schechter:1980gr}
J.~Schechter and J.~W.~Valle,
Phys.\ Rev.\ D {\bf 22} (1980) 2227;
R.~N.~Mohapatra and G.~Senjanovic,
Phys.\ Rev.\ D {\bf 23} (1981) 165.

\bibitem{Maltoni:2002ni}
M.~Maltoni, T.~Schwetz, M.~A.~Tortola and J.~W.~Valle,
Phys.\ Rev.\ D {\bf 67} (2003) 013011.

\bibitem{Gonzalez-Garcia:2003qf}
M.~C.~Gonzalez-Garcia and C.~Pena-Garay,
arXiv:hep-ph/0306001;
P.~C.~de Holanda and A.~Y.~Smirnov,
JCAP {\bf 0302} (2003) 001
[arXiv:hep-ph/0212270];
A.~B.~Balantekin and H.~Yuksel,
J.\ Phys.\ G {\bf 29} (2003) 665;
H.~Nunokawa, W.~J.~Teves and R.~Zukanovich Funchal,
Phys.\ Lett.\ B {\bf 562} (2003) 28.

\bibitem{Fukugita:1986hr}
M.~Fukugita and T.~Yanagida,
Phys.\ Lett.\ B {\bf 174}, 45 (1986).

\bibitem{Affleck:1984fy}
I.~Affleck and M.~Dine,
Nucl.\ Phys.\ B {\bf 249}, 361 (1985);
M.~Dine, L.~Randall and S.~Thomas,
Phys.\ Rev.\ Lett.\  {\bf 75}, 398 (1995);
Nucl.\ Phys.\ B {\bf 458}, 291 (1996).

\bibitem{Fukugita:1990gb}
M.~Fukugita and T.~Yanagida,
Phys.\ Rev.\ D {\bf 42}, 1285 (1990);
A.~E.~Nelson and S.~M.~Barr,
Phys.\ Lett.\ B {\bf 246}, 141 (1990);
J.~A.~Harvey and M.~S.~Turner,
Phys.\ Rev.\ D {\bf 42}, 3344 (1990);
W.~Fischler, G.~F.~Giudice, R.~G.~Leigh and S.~Paban,
Phys.\ Lett.\ B {\bf 258}, 45 (1991).

\bibitem{Campbell:1990fa}
B.~A.~Campbell, S.~Davidson, J.~R.~Ellis and K.~A.~Olive,
Phys.\ Lett.\ B {\bf 256}, 457 (1991);
B.~A.~Campbell, S.~Davidson, J.~R.~Ellis and K.~A.~Olive,
Astropart.\ Phys.\  {\bf 1}, 77 (1992).

\bibitem{Dreiner:vm}
H.~K.~Dreiner and G.~G.~Ross,
Nucl.\ Phys.\ B {\bf 410}, 188 (1993).

\bibitem{Davidson:1997mc}
S.~Davidson and J.~R.~Ellis,
Phys.\ Rev.\ D {\bf 56}, 4182 (1997);
S.~Davidson,
arXiv:hep-ph/9808427.

\bibitem{Kuzmin:1985mm}
V.~A.~Kuzmin, V.~A.~Rubakov and M.~E.~Shaposhnikov,
Phys.\ Lett.\ B {\bf 155}, 36 (1985).




\end{thebibliography}
\end{document}